\documentclass[aps,pre,showpacs,twocolumn,superscriptaddress]{revtex4}

\font\mb=msbm10

\begin{document}

\newcommand{\vl}{\vec{l}}
\newcommand{\mt}{\mu_{t}}
\newcommand{\md}{{\bf{M}}^{-\tau}}
\newcommand{\ds}{\Delta_{\rm i}^{\tau}S}
\newcommand{\sk}{s_{k}}
\newcommand{\vk}{\vec{k}}
\newcommand{\G}{X}
\newcommand{\vrr}{\vec{r}}
\newcommand{\cM}{{\cal{M}}}
\newcommand{\vL}{\vec{L}\left[{\Phi}^{-t}(\vl,X,0)\right]}
\newcommand{\vv}{\vec{v}}
\newcommand{\cb}{\cal{B}}
\newcommand{\re}{\rho_{\rm eq}}
\newcommand{\intk}{{\frac{1}{|\cal{B}|}}\int_{\cal{B}}\,d\vec{k}\;F_{\vk}}
\newcommand{\intknu}{{\frac{\nu(\cM)}{|\cal{B}|}}\int_{\cal{B}}\,d\vec{k}\;
F_{\vk}}
\newcommand{\dlt}{\vec{d}(X,t)}
\newcommand{\A}{\{ A_i \}}
\newcommand{\ct}{{\cal{T}}}

\newtheorem{definition}{Definition}
\newtheorem{lemma}{Lemma}[definition]


\title{Entropy production of diffusion in spatially periodic deterministic
systems}

\author{J. R. Dorfman}
\affiliation{Department of Physics and Institute for Physical Science and
Technology, University of Maryland, College Park, MD 20742, USA}
\email{jrd@ipst.umd.edu}
\author{P. Gaspard}
\affiliation{Center for Nonlinear Phenomena and Complex Systems,
Universit\'e Libre de Bruxelles, Code Postal 231, Campus Plaine, B-1050
Brussels, Belgium}
\email{gaspard@ulb.ac.be}
\author{T. Gilbert}
\affiliation{Department of Chemical Physics,
The Weizmann Institute of Science, Rehovot 76100, Israel}
\email{thomas.gilbert@weizmann.ac.il}
\date{\today}


\begin{abstract}

This paper presents an {\it ab initio} derivation of the expression
given by irreversible thermodynamics for the rate of entropy production
for different classes of diffusive processes. The first class are Lorentz
gases, where non-interacting particles move on a spatially periodic
lattice, and collide elastically with fixed scatterers. The second
class are 
periodic systems where $N$ particles interact with each other, and one of
them is a
tracer particle which diffuses among the cells of the lattice. 
We assume that, in either
case, the dynamics of the system is deterministic and hyperbolic,
with positive Lyapunov exponents. This work extends
methods originally developed for a chaotic
two-dimensional model of diffusion, the multi-baker map, to higher
dimensional, continuous time dynamical
systems appropriate for systems with one or more moving
particles. Here we express the rate of entropy
production in terms of hydrodynamic measures that are determined by the
fractal properties of microscopic
hydrodynamic modes that describe the slowest decay of the system
to an equilibrium state.
\end{abstract}


\maketitle


\section{Introduction}


In 1902, Gibbs described a mechanism by which the entropy could increase toward
its equilibrium value in mechanical systems obeying Newton's equations
\cite{Gibbs}.  Gibbs' mechanism is based on the assumption that the
microscopic dynamics is mixing.  The mixing would allow coarse-grained
probabilities to reach their equilibrium values after a long time,
a result that has received a rigorous meaning in the modern definition of
mixing
\cite{Do99}.  The second
ingredient of Gibbs' mechanism is the assumption that the entropy of a physical
system should be defined as a quantity which is now referred to as the
coarse-grained entropy.
The use of the coarse-grained entropy could be
justified by the fact that, if the entropy should be given according to
Boltzmann by the logarithm of the number of complexion of a system,
then it can only
be defined by introducing cells of non-vanishing size in systems described by
continuous coordinates.\footnote{With the advent of quantum mechanics this size
turned out
to be equal to $h^f$ where $h$ is the Planck constant and $f$ the number of
degrees of freedom.  In classical systems, the size of the cells remains
arbitrary, which is related to the fact that classical physics does not fix the
constant of entropy.}

The aim of the present paper is to apply
the program set up by Gibbs to hyperbolic, deterministic dynamical systems
sustaining a
transport process of diffusion.  We suppose that the systems obey Liouville's
theorem, namely that phase space volumes are preserved by the dynamics,
which is a major
assumption used by Gibbs. Specifically, the systems we consider are either
periodic Lorentz gases, in which moving particles diffuse through a
lattice interacting only with fixed scatterers, or are periodic
repetitions of interacting $N$-particle systems such that a tagged particle
is followed as it undergoes diffusion among the unit cells. The present work is
an extension to continuous time, and interacting systems, of previous
works \cite{Ga97,TaGa99,TaGa00,GiDoGa00,GiDo00} concerned with the
multi-baker map,
as a chaotic model of diffusion. In the previous work, we used the fact that
an initial non-equilibrium distribution function, in the multi-baker
``phase space'',
rapidly develops a fractal structure due to the chaotic nature of the
dynamics. This structure is such that variations of the
distribution function on arbitrarily fine scales develop as the
system evolves in time. The final stages of the approach to equilibrium are
then
controlled by the decay of fractal, microscopic hydrodynamic modes of the
system, in
this case, diffusive modes, which decay with time as $\exp(-Dk^2t)$,
where $k$ is a wave number characterizing a particular mode associated
with a macroscopic density variation over a distance of order $k^{-1}$, $D$ is
the diffusion coefficient, and $t$ is the time. For the multi-baker
system it is possible to express the rate of entropy production in this
final stage in terms of measures of sets which are determined by the
non-equilibrium phase space distribution in the set, in particular, by
the values of the fractal hydrodynamic modes in the set.

In this paper we show that it is possible to apply the same methods to
 calculate the rate of entropy production for diffusive flows in
 periodic Lorentz gases and for tracer diffusion in periodic, interacting
 $N$-particle systems, as long as the microscopic dynamics is
 deterministic, mixing, and chaotic.
Our method is based on the explicit construction of the microscopic fractal
hydrodynamic modes of
diffusion, which characterize the long time relaxation of the system toward
thermodynamic
equilibrium. Our main result is that we obtain by this method exactly the
expression for the
 rate of entropy production as given by irreversible thermodynamics
 for these systems\cite{deGMaz}. The source of this agreement can be traced
to the
role played by the fractal hydrodynamic modes, both for requiring a
 coarse graining of the phase space to properly incorporate the
 effects of their fractal properties on entropy production in the system,
as well
 as for describing the slowest decay of the system as it relaxes to
equilibrium.

The plan of the paper is the following. The mathematical methods
needed to describe spatially periodic systems are provided in
Sec. \ref{PerSys}.  Then the non-equilibrium distribution is defined in
Sec. \ref{Non-Equil}. Once the distribution function has been constructed,
we can identify the microscopic hydrodynamic modes of diffusion, and they
are then constructed in Sec. \ref{Hydro}. Here
we use the properties of these modes to introduce the notion of a
hydrodynamic measure of a set in phase space, and we also identify the
sets that are used in the coarse graining of the phase space. These
sets are not arbitrary, but must have some specific properties in
order to be useful for the calculation of the rate of entropy
production, which
is carried out in Sec. \ref{EntrProd}. We conclude with a
discussion of the method and results obtained here, and with an
outline of directions for future work in Sec. \ref{Concl}.


\section{Spatially periodic systems}

\label{PerSys}

\subsection{The phase space}

We consider a deterministic dynamical system of phase space dimension $M$
which is
spatially periodic in the form of a $d$-dimensional lattice ${\cal L}$.  We
will label the positions of the periodic cells $\cM$ on the lattice by the
vector $\vl \in{\cal L}$, and the phase space coordinates within an 
elementary cell by
$\G\in\cM$.  The lattice $\cal L$ is isomorphic to $\mbox{\mb Z}^d$.  The
total phase
space of the system is the direct product $\cM\otimes{\cal L}$ of dimension
$M$.
The time displacement operator over a time  interval $t$, acting on points
$(\vl,
\G)$, is denoted $\Phi^{t}$, which is also called the {\it flow}.  On the other
hand, the time displacement operator acting on points $\G$ inside the basic
unit
cell is denoted $\phi^{t}$.

Examples of such systems are the following.

\subsubsection{The two-dimensional Lorentz gas}

In this system a point particle moves in
free flight and undergoes elastic collisions on hard disks forming a triangular
lattice.  The phase space is defined by the spatial and velocity
coordinates $(\vrr,\vv)$ of the moving particle.  The spatial coordinates
vary over
the plane with the exclusion of the area occupied by the disks:
$\vrr\in{\cal Q}$
with ${\rm dim}\:{\cal Q}=2$.  The velocity coordinates form another
two-dimensional
vector $\vv\in{\mbox{\mb R}}^2$.  Energy is conserved during the motion so
that
each energy shell is preserved by the dynamics.  In each energy shell, we
can thus
reduce the coordinates to the two positions $\vrr=(x,y)\in{\cal Q}$ and the
velocity angle $\varphi \in[0,2\pi[$.  The position space decomposes into a
triangular lattice of hexagonal elementary cells, each containing a single
disk:
${\cal
Q}=\cup_{\vl\in{\cal L}} {\cal T}^{\vl}{\cal C}$, where ${\cal T}^{\vl}$
denotes the translation by the lattice vector $\vl$, and ${\cal{C}}$ is
  the elementary cell of the position space ${\cal{Q}}$.  The
elementary cell
of the phase space has thus the coordinates
$\G=(x,y,\varphi)\in\cM={\cal C}\otimes[0,2\pi[$.  In the two-dimensional
Lorentz
gas, the phase space has the dimension ${\rm dim}\:{\cal M}=3$ while the
lattice has
the dimension $d={\rm dim}\:{\cal L}=2$.  The flow of the hard-disk Lorentz gas
preserves the Lebesgue measure $d\G=dx\, dy\, d\varphi$.  We notice that the
horizon of the hard-disk Lorentz gas must be finite in order for the diffusion
coefficient to be finite \cite{BuSi80}, which we assume in the following.

Similar considerations apply to the Lorentz gas in which a point particle
moves in a
periodic lattice of attractive Yukawa potentials. In this system the
diffusion coefficient is
positive and finite for a large enough energy \cite{Kn87}. Here, too,
the phase space has $3$ dimensions and is the union of the constant
energy surfaces for each cell of the lattice.

\subsubsection{The three-dimensional Lorentz gas}

This Lorentz gas is the direct generalization of the
two-dimensional one.  The spatial coordinates vary over the space with the
exclusion
of the volume occupied by the spheres:
$\vrr\in{\cal Q}$ with ${\rm dim}\:{\cal Q}=3$.  The velocity coordinates form
another three-dimensional vector $\vv\in{\mbox{\mb R}}^3$.  In each
energy shell, we can reduce the coordinates to the three positions
$\vrr=(x,y,z)\in{\cal Q}$ and the two velocity spherical angles
$\theta\in[0,\pi]$ and $\varphi\in[0,2\pi[$.  The position space decomposes
into a
lattice of elementary cells containing one or several disks: ${\cal
Q}=\cup_{\vl\in{\cal L}} {\cal T}^{\vl}{\cal C}$.  The elementary cell of
the phase space has thus the coordinates
$\G=(x,y,z,\cos\theta,\varphi)\in\cM={\cal
  C}\otimes[-1,+1]\otimes[0,2\pi[$, where, again, ${\cal{C}}$ is
  the elementary cell of the position space ${\cal{Q}}$.
In the
three-dimensional Lorentz gas, the phase space has the dimension ${\rm
dim}\,{\cal
M}=5$ while the lattice has the dimension $d={\rm dim}\:{\cal L}=3$.  The
flow of
the hard-disk Lorentz gas preserves the Lebesgue measure $d\G=dx\, dy\,dz\,
d\cos\theta\, d\varphi$.  Here also, we suppose that the horizon is finite
in order
for the diffusion coefficient to be finite.

\subsubsection{Diffusion of a tracer in a system on a torus}

The molecular dynamics simulation of the diffusion of a tracer particle
moving in
a fluid can be performed by considering a finite number of particles
modeling the fluid and the tracer particle, all of them moving with
interactions, in a domain
delimited by periodic boundary conditions.  The total number of particles is
equal to $N$.  The center of mass can be taken at rest.  The vector
$\vl$
can be used to locate the position of the cell containing the tracer
particle as it moves on the
checkerboard lattice made of
infinitely many images of the system, which tiles the $d$-dimensional
space of the system. Then the diffusion
coefficient
of the tracer particle can be computed by adding the appropriate
lattice
vector to $\vl$ each time the tracer particle crosses a boundary.  The
energy
and total momentum are to be conserved so
that here $\G$ denotes the phase space coordinates of an elementary cell of the
phase space, which is of dimension $M={\rm dim}\:{\cal M}=2dN-2d-1$ after
elimination
of the $d$ degrees of freedom of the center of mass, while the
lattice is of dimension $d={\rm dim}\:{\cal L}=2$ or $3$.

In summary, we will denote by $\G$ the coordinates on an
energy-momentum shell of a
micro-canonical ensemble for the periodic system.  We suppose that the flow
$\Phi^t$ preserves the Lebesgue measure $d\G$.  Moreover, we assume
that the diffusion coefficient of the system is finite, that the mean drift
vanishes, and that the microscopic dynamics is chaotic and mixing. In the
sequel $\vl$ will denote the lattice position vector and
${\cal T}$ the translation operator on the lattice. Subsets of the unit
cell $\cM$ will be denoted by capital Roman letters.
The notation $A$ will be used for a set belonging to the elementary phase
space cell
at the origin $\vl=0$ of the lattice. The notation $A_{\vl}={\cal
T}^{\vl}A$ will be used
when we want to refer to a set in a specific lattice cell at
position $\vl$.

\subsection{Lattice Fourier transforms}

We will also need to define lattice Fourier
transforms \cite{Ga98}. We will need the
preliminary result~:
\begin{definition}\label{def1}
Consider a function $G(\vl,\G)$, which
is a function of the lattice coordinate $\vl$, and the unit cell
coordinate $\G$, then this function can be expressed in terms of a
{\em lattice Fourier transform} as
\begin{equation}
G(\vl,\G) = \frac{1}{|\cb|}\int_{\cb}d\vk
\; e^{i\vk\cdot\vl}\; \tilde G(\vk,\G)\ ,
\label{1}
\end{equation}
where $\cb$ denotes the first Brillouin zone of the reciprocal
lattice, $|\cb|$ is its volume, and
$\tilde G$ is the lattice Fourier transform of $G$.
\end{definition}
It is important to note that~:
\begin{lemma}
If $G$ is only a
function of the lattice vector $\vl$, then $\tilde G$ does not depend
upon the unit cell coordinate $\G$, but only upon $\vk$.
\end{lemma}


\section{The non-equilibrium distribution}

\label{Non-Equil}

We consider a periodic deterministic dynamical system with finite diffusion
coefficient.


\subsection{The non-equilibrium measure}


We now construct the statistical ensemble we will use for the
rest of this paper.  We suppose that the coordinates of the ensemble are
distributed on the lattice in such a way that the distribution can be
described by
an initial ensemble density $\rho(\vl,\G,0)$, where $\rho$ denotes the
number of systems per unit phase space volume. We take this density to
be close to that of total equilibrium and write it in the form
\begin{equation}
\rho(\vl,\G,0) =\re\; \left[1+R(\vl,\G,0)\right]\ ,
\label{2}
\end{equation}
where the equilibrium distribution $\re$ is, for Lorentz gases, uniform
with respect to
the cells $\vl$, and with respect to the phase variables $\G$ in
agreement with
the assumption that the Lebesgue measure $d\G$ is preserved by the flow
$\Phi^t$. For the case of tracer diffusion, $\re$ is the equilibrium
micro-canonical distribution for the $N$ particles in a cell on the lattice.
The initial deviation from equilibrium in the cell located at
$\vl$ is denoted by $R(\vl,\G,0)$, which we assume to be Lebesgue
integrable when weighted with the equilibrium distribution.
Using the lattice Fourier transform, Eq. (\ref{1}),
we can express this deviation in the form
\begin{equation}
R(\vl,\G,0) = {\frac{1}{|\cal{B}|}}\int_{\cal{B}}\,d\vec{k}\;
e^{i\vk\cdot\vl}\; \tilde R(\vk,\G)\ .
\label{3}
\end{equation}
The phase space density (\ref{2}) leads us to the definition of the
non-equilibrium measure of a set $A_{\vl}$ belonging to the phase space cell
${\cal T}^{\vl}\cM$ corresponding to the lattice vector $\vl$~:
\begin{definition}
\label{def:measures}
The equilibrium measure $\nu(A_{\vl})$ of a set $A_{\vl}$ is defined by
\begin{equation}
\nu(A_{\vl}) \equiv \int_{A_{\vl}} d\G \re\ ,
\label{12}
\end{equation}
and the non-equilibrium measure, $\mt(A_{\vl})$, of the same set by
\begin{equation}
\mt(A_{\vl}) \equiv \int_{A_{\vl}} d\G \rho(\vl,\G,t) =\nu(A_{\vl})
+\delta\mt(A_{\vl})\ ,
\label{13}
\end{equation}
where
\begin{equation}
\delta\mt(A_{\vl}) = \int_{A_{\vl}}d\G \; \re R(\vl,\G,t)\; .
\end{equation}
\end{definition}

We will simplify matters a bit by assuming that the initial deviation
from equilibrium $R(\vl,\G,0)$ depends only upon the cell $\vl$ but
not on the initial phase $\G$ of the system within the cell.
In this case the Fourier transform $\tilde R(\vk,\G)$ does not
depend upon $\G$, either, and will henceforth be denoted by
$F_{\vk}$.

The time dependent distribution function
$\rho(\vl,\G,t)$ is the solution of Liouville's equation and is given
by
\begin{eqnarray}
\rho(\vl,\G,t) &=& \re\left[ 1 +R(\vl,\G,t)\right]\ ,\label{4}\\
&=&\re\left\{ 1+
{\frac{1}{|\cal{B}|}}\int_{\cal{B}}\,d\vec{k}\; F_{\vk} \;
e^{i\vk\cdot\vL}\right\}\ .\nonumber
\end{eqnarray}
Here $\vec{L}$ denotes the projection on the lattice
coordinate, thus $\vL$ is the lattice vector of the cell in which a moving
particle would be located at time $-t$ if it were in cell $\vl$ at
time $t=0$, with phase $\G$.
We can express the time dependent
deviation from total equilibrium in the form
\begin{equation}
R(\vl,\G,t) =\intk \; e^{i\vk\cdot\left[\vl+\dlt\right]}\ ,
\label{5}
\end{equation}
where the {\em backward} displacement of the lattice vector $\vl$ over a time
interval $t$ is defined by
\begin{equation}
\dlt = \vL -\vl\ .
\label{6}
\end{equation}
We will make heavy use of the fact that $\dlt$ depends upon the time
interval $(-t,0)$ and upon the phase point $\G$, at the initial time,
but not upon the initial cell $\vl$. In other words, the periodicity of
the lattice and the dynamics produce a ``winding
number'' $\dlt$ that does not depend upon the cell in which the
trajectory is located at the initial time. We remark here that the long
time limit of the displacement vector $\dlt$
may be a wildly varying function of the phase coordinate $\G$. Thus we
expect that the decaying modes of the time
dependent distribution (\ref{4})
are singular functions of the phase coordinates.

We define the time dependent density of the tracer particle by integrating the
phase space density over the coordinates $\G$ of an elementary phase space
cell of
the lattice \footnote{We will use the term {\it density of the tracer
particle} to refer both to the density
of moving particles in a Lorentz gas as well as to the density of the
tracer particle in the system of $N$ moving, interacting particles.} :
\begin{equation}
n(\vl,t) \equiv  \int_{\cM} d\G \; \rho(\vl,\G,t) = \mu_t(\cM_{\vl})\ .
\label{6/7}
\end{equation}

Using Eq. (\ref{5}), we obtain the density as
\begin{eqnarray}
\lefteqn{n(\vl,t) = \nu(\cM)}&&\label{7}\\
&&\times\left[ 1 + \intk
\; e^{i\vk\cdot\vl} \; \frac{\int_{\cM} d\G \re
\; e^{i\vk\cdot\dlt}}{\int_{\cM} d\G\,\re} \right]\ .
\nonumber
\end{eqnarray}

We consider times $t$ that are long compared to the mean time between
collisions of
the moving particles, but short compared to the time needed for the system to
relax to total equilibrium. For such times, we expect the time dependent
deviation from total equilibrium to decay exponentially with a rate $-s_k$
given
by the van Hove relation \cite{vH54}
\begin{eqnarray}
s_k &\equiv& \lim_{t\to\infty} \frac{1}{t} \ln \langle
e^{i\vk\cdot\dlt}\rangle_{\cM}\ ,\nonumber\\
&=& \lim_{t\to\infty} \frac{1}{t} \ln \frac{\int_{\cM} d\G \re
\; e^{i\vk\cdot\dlt}}{\int_{\cM} d\G\re}\ ,
\label{7/9}
\end{eqnarray}
which gives the decay rate of a hydrodynamic mode of diffusion of wave number
$\vk$.  An expansion in powers of the wave number gives
\begin{equation}
\sk = -D \vk^2 + O(\vk^4)\ ,
\label{9}
\end{equation}
with diffusion coefficient $D$.  We have here assumed that the diffusive motion
of the tracer particle is invariant under space inversion so that all the odd
powers of the wave number vanish.  We notice that the existence of the
successive
terms of the expansion in powers of the wave number depends on the existence of
the super-Burnett and higher diffusion coefficients\cite{HVB}, which has
been recently
proved for the hard-disk periodic Lorentz gas with a finite horizon
\cite{ChDe00}.

The definition (\ref{7/9}) shows that
\begin{equation}
\frac{\int_{\cM} d\G \re \; e^{i\vk\cdot\dlt}}{\int_{\cM} d\G \re} =
C(\vk,t) \;
e^{s_kt}\ ,
\label{8}
\end{equation}
where $C(\vk,t)$ is a function of the wave number $\vk$ with a sub-exponential
dependence on time, i.~e.
\begin{equation}
\lim_{t\to\infty} \frac{1}{t} \ln C(\vk,t) = 0\ .
\label{subexp}
\end{equation}
Accordingly, the tracer density can be written as
\begin{equation}
n(\vl,t) = \nu(\cM)\; \left[ 1 + \intk
\; e^{i\vk\cdot\vl} \; C(\vk,t) \; e^{s_kt} \right]
\label{8/7}
\end{equation}
Notice that $n(\vl,t)$ obeys a form of the diffusion equation,
appropriate for our lattice system, given by
\begin{eqnarray}
\frac{\partial n(\vl,t)}{\partial t} &=& \intknu \, e^{i\vk\cdot\vl}
\, C(\vk,t) \, \sk \, e^{\sk t} +\cdots  \nonumber  \\
&=& -D\; \intknu \, \vk^{2} \, e^{i\vk\cdot\vl}
C(\vk,t) \, e^{\sk t}\nonumber\\
&+&\cdots\ ,\label{10}
\end{eqnarray}
which, in the scaling limit where the size of the unit cell becomes
small and for large times and small wave numbers, is
the diffusion equation:
\begin{equation}
\frac{\partial n(\vl,t)}{\partial t} \simeq D \;
\frac{\partial^2 n(\vl,t)}{\partial \vl^2}\ .
\label{diff}
\end{equation}


\section{The hydrodynamic modes}

\label{Hydro}


\subsection{The hydrodynamic measures}


Spatially periodic deviations from total equilibrium characterized by the wave
number $\vk$ relax exponentially at the rate given by van Hove's relation
\cite{vH54}.  Our purpose is here to determine the non-equilibrium state
corresponding to
this mode of exponential relaxation.  This state can be defined as a
measure, which we call a  hydrodynamic
measure, associated with the hydrodynamic mode of diffusion. It is the
microscopic analog
of the solutions $\exp(i{\vk}\cdot{\vl}- D k^2 t)$ of wave number $\vk$ for the
macroscopic diffusion equation (\ref{diff}).

We introduce the hydrodynamic measures by considering the deviations from the
equilibrium measure for a set $A_{\vl}$:
\begin{eqnarray}
\delta\mt(A_{\vl}) &=&  \int_{A_{\vl}}d\G \;\re R(\vl,\G,t)\ , \nonumber \\
&=&  \int_{A_{\vl}}d\G \re \intk \; e^{i\vk\cdot[\vl+\dlt]}\ ,  \nonumber
\\
&=& \intknu \; e^{i\vk\cdot\vl} \; \frac{\int_{\cM}d\G\ \re \;
e^{i\vk\cdot\dlt}}{\int_{\cM}d\G\re} \nonumber\\
&&\times \frac{\int_{A}d\G \re \;
e^{i\vk\cdot\dlt}}{\int_{\cM}d\G \re
\; e^{i\vk\cdot\dlt}}\ ,
\label{14}
\end{eqnarray}
where we have used the property that the backward displacement $\dlt$ is
independent of the initial lattice vector $\vl$ so that the integral over
the set
$A_{\vl}$ is equal to the integral over the set $A$ of the
elementary cell at the
origin
of the lattice.  In the last line, we have factorized the exponential decay
according to Eq. (\ref{8}), which is independent of the set $A$, from
a further factor,
which depends on the set $A$ but which is expected to have a well-defined limit
for $t\to\infty$ because both its numerator and denominator are expected to
decay
exponentially as $\exp(s_kt)$.  This observation motivates the
\begin{definition}
\label{def:hyme}
The hydrodynamic measure $\chi_{\vk}(A,t)$ is defined by
\begin{equation}
\chi_{\vk}(A,t) \equiv \nu(\cM)\; \frac{\int_{A}d\G \re \;
e^{i\vk\cdot\dlt}}{\int_{\cM}d\G \re
\; e^{i\vk\cdot\dlt}}\ .
\label{17}
\end{equation}
\end{definition}
We emphasize that the hydrodynamic measures are independent of the cell
location $\vl$.  We notice that the hydrodynamic measures are complex measures
because of the lattice Fourier transform.

One important property of the hydrodynamic measures is that the total
hydrodynamic
measure of a unit cell is constant in time, as follows from the definition
\ref{def:hyme}.  That is,
\begin{lemma}
For the set $A = \cM$, i.~e., the phase space region associated with an
entire unit
cell, the hydrodynamic measure is
\begin{equation}
\chi_{\vk}(\cM,t) = \nu(\cM)\; .
\label{19}
\end{equation}
\end{lemma}
Another observation is that
\begin{lemma}
If we make a $\vk$-expansion of the hydrodynamic measure
of the form
\begin{equation}
\chi_{\vk}(A,t) = \nu(A) + i\vk\cdot\vec{T}(A,t) + \vk\vk:{\tensor g
}(A,t)
+\cdots
\; ,
\label{20}
\end{equation}
it follows from Eq. (\ref{19}), that $\vec{T}(\cM,t) =0$ and that
$\tensor g(\cM,t)=0$, etc.
\end{lemma}
Properties like these have already been used
in the various symmetric, multi-baker models
\cite{Ga97,TaGa99,TaGa00,GiDoGa00}.  For a system with vanishing mean
drift $\langle \dlt\rangle_{\cM}=0$, the two first coefficients of the
$\vk$-expansion of the hydrodynamic measures can be expressed as
\begin{eqnarray}
\vec{T}(A,t) &=& \nu(A)\; \langle \dlt\rangle_A \\
\tensor g
(A,t) &=& \frac{1}{2}\; \nu(A) \\
&&\times\left[ \; \langle \dlt
\dlt\rangle_{\cM} -\; \langle \dlt\dlt\rangle_A\right]
\nonumber\end{eqnarray}
with the definition
\begin{equation}
\langle \cdot\rangle_A \equiv \frac{ \int_Ad\G \re (\cdot)}{\int_A d\G\re}
\end{equation}

Thanks to the hydrodynamic measures Eq. (\ref{17}) and Eq. (\ref{8}), we
finally
derive
from Eq. (\ref{14}) an expression for the measure of a set $A$ in cell
$\cM_{\vl}$~:
\begin{eqnarray}
\mt(A_{\vl}) &=& \nu(A) + \delta\mt(A_{\vl})\ ,\nonumber\\
 &=& \nu(A) \label{18}\\
&&+ \intk \;
e^{i\vk\cdot\vl}\;
C(\vk,t)\; e^{\sk t}\; \chi_{\vk}(A,t)\; .
\nonumber
\end{eqnarray}


\subsection{Conservation of measure and de Rham-type equation}


Since the time evolution is a measure preserving Liouville operator,
the measure of any set $A$ remains constant as the set follows the
motion of the system in phase space. Therefore we may express this
conservation of measure as
\begin{equation}
\mu_{t+\tau}(A) = \mt(\Phi^{-\tau}A).
\label{21}
\end{equation}
Here $\tau$ denotes some time interval, and $\Phi^{-\tau}A$ is the
pre-image of the set $A$ under the flow, obtained by following the backward
evolution of the points of $A$ over a time interval $\tau$.
This simple result has some important consequences, among them, a de Rham-type
equation for the hydrodynamic measures.

We suppose that the sets $A$ are sufficiently small that all the
points in them will flow through the same sequence of cells over some large
time interval $(-\ct \leq \tau \leq \ct)$.
In such a case the set of points $\Phi^{-\tau}A$ are all in
the {\it same} cell with location denoted by $\vl +\vec{d}(\G_A,\tau)$ which is
determined by the backward evolution of an arbitrary phase point $\G_A$ in
the set $A$. Using Eq. (\ref{18}), we can express the application of Eq.
(\ref{21}) to a set
$A_{\vl}$ as
\begin{widetext}
\begin{equation}
\intk \; e^{i\vk\cdot\vl}\; C(\vk,t+\tau)\; e^{(t+\tau)\sk}\;
\chi_{\vk}(A,t+\tau)
=
\intk \; e^{i\vk\cdot[\vl +\vec{d}(\G_A,\tau)]}\; C(\vk,t)\; e^{\sk
t}\; \chi_{\vk}(\Phi^{-\tau}A,t)\; .
\label{22}
\end{equation}
\end{widetext}
Since this equation must be true for all cells $\vl$ for all $F_{\vk}$,
and for all sets $A$ satisfying the above condition, the only way it can be
satisfied is if the integrands are equal almost everywhere. Equating the
integrands
leads to the equation
\begin{eqnarray}
\lefteqn{C(\vk,t+\tau)\;e^{\sk\tau} \;  \chi_{\vk}(A,t+\tau) =}&&\nonumber\\
&&e^{i\vk\cdot\vec{d}(\G_A,\tau)}\; C(\vk,t)\; \chi_{\vk}(\Phi^{-\tau}A,t)\; .
\label{23}
\end{eqnarray}
Since the hydrodynamic measures do not depend on the lattice vector, the
pre-image $\Phi^{-\tau}A$ under the full flow over the lattice can be
reduced to
the pre-image $\phi^{-\tau}A$ under the flow defined with periodic boundary
conditions inside the elementary cell $\cM$ at $\vl=0$.

Now we suppose that the dynamics is hyperbolic in order to assert that the
hydrodynamic measures $\chi_{\vk}(A,t)$ reach asymptotic forms exponentially
rapidly, that is, on a time scale of the order of the inverse of the positive
Lyapunov exponent for the system. In this case, we can replace the hydrodynamic
measures in Eq. (\ref{23}) by their asymptotic forms, denoted by
$\chi_{\vk}(A)$,
in the long time limit $t\to\infty$.  Moreover, we have the property that
\begin{equation}
\lim_{t\to\infty} \; \frac{C(\vk,t+\tau)}{C(\vk;t)}=1
\label{ratioC}
\end{equation}
as a consequence of the sub-exponential behavior (\ref{subexp}) of the
functions
$C(\vk,t)$.

In the long
time limit $t\to\infty$, combining Eqs. (\ref{23}) and
(\ref{ratioC}), we obtain the
\begin{lemma}
\label{lem:derham}
The hydrodynamic measures satisfy a de Rham-type equation~:
\begin{equation}
e^{\sk\tau} \; \chi_{\vk}(A) =
e^{i\vk\cdot\vec{d}(\G_A,\tau)}\; \chi_{\vk}(\phi^{-\tau}A)\; .
\label{23a}
\end{equation}
\end{lemma}
Explicit solutions of this equation have been found for multi-baker
maps \cite{GiDo00,TaGa95,GiDoGa01}.
For the hard-disk Lorentz gas, these solutions lead to the cumulative
functions constructed in reference \cite{GaClGiDo01} with one-dimensional sets
$A$ and in reference \cite{Ga98} with two-dimensional sets $A$.  An alternative
form of Eq. (\ref{23a}) is
\begin{equation}
e^{\sk\tau} \; \chi_{\vk}(\phi^{\tau}A) =
e^{i\vk\cdot\vec{d}(\G_{\phi^{\tau}A},\tau)}\; \chi_{\vk}(A)\; .
\label{23b}
\end{equation}
Equation (\ref{23b}) has an expansion in powers of the
wave\-number $\vk$ which will be useful in the calculation of the rate
of entropy production. In obtaining these expansions we will make use
of the $\vk$-expansions of the $\chi_{\vk}$ functions given in Eq.
(\ref{20}).
The wave number expansion of Eq. (\ref{23b}) leads to the following
equation for terms of order $\vk$:
\begin{equation}
\vk\cdot\vec{T}(\phi^{\tau}A) = \vk\cdot
\vec{T}(A) +\nu(A)\; \vk\cdot\vec{d}(\G_{\phi^{\tau}A},\tau) \; .
\label{29}
\end{equation}


\subsection{Partition of phase space and sum rules}


We consider a partition $\{A_j\}$ of the elementary cell $\cM$ at $\vl=0$ of
the phase space into disjoint sets $A_j$:
\begin{equation}
\cup_{A_{j}\subset \cM}\; A_{j} = \cM\ ,\quad A_i \cap A_j = \emptyset,\;
\forall\; i,\;j,\;i\neq j\ .
\label{pre11}
\end{equation}
We notice that the images $\phi^{\tau}A_j$ also form a partition of the
elementary cell of phase space:
\begin{equation}
\cM =\cup_{A_{j}\subset \cM}\; \phi^{\tau}A_{j}\; .
\label{pre11bis}
\end{equation}

We can apply the de Rham-type equation (\ref{23b}) to one set $A_j$ of the
partition (\ref{pre11}) and sum both members of Eq. (\ref{23b}) over all
the sets
$A_j\in\cM$ to obtain:
\begin{equation}
e^{\sk\tau} \; \sum_j\, \chi_{\vk}(\phi^{\tau}A_j) = \sum_j\;
e^{i\vk\cdot\vec{d}_j}\; \chi_{\vk}(A_j)\; ,
\label{23c}
\end{equation}
with the notation
\begin{equation}
\vec{d}_j \equiv \vec{d}(\G_{\phi^{\tau}A_j},\tau) \; .
\end{equation}
Since the sets $\phi^{\tau}A_j$ form a partition of
$\cM$ into disjoint sets, we infer from Eq. (\ref{19}) that
\begin{equation}
\sum_j\, \chi_{\vk}(\phi^{\tau}A_j) = \chi_{\vk}(\cM) = \nu(\cM) \; .
\label{sum1}
\end{equation}
so that Eq. (\ref{23c}) becomes
\begin{equation}
\nu(\cM)\; e^{\sk\tau} = \sum_j\;
e^{i\vk\cdot\vec{d}_j}\; \chi_{\vk}(A_j)\; .
\label{23d}
\end{equation}

Now, we perform a wavenumber expansion of both members of Eq.
(\ref{23d}) using
Eqs. (\ref{9}) and (\ref{20}).  Using the properties
\begin{eqnarray}
\sum_j \nu(A_j) &=& \nu(\cM)\ , \\
\sum_j \vec{T}(A_j) &=& \vec{T}(\cM) = 0 \label{sumT=0}\ ,\\
\sum_j \tensor g
(A_j) &=& \tensor g
(\cM) = 0 \label{sumg=0}\ ,
\end{eqnarray}
the identification of the terms which are of the first and second powers of the
wave number $\vk$ gives us the following two sum rules:
\begin{eqnarray}
&&\sum_j \vec{d}_j \; \nu(A_j)= 0\ , \\
&&\sum_j \left[ \vec{d}_j \; \vec{T}(A_j) + \vec{T}(A_j) \; \vec{d}_j +
\vec{d}_j \; \vec{d}_j \; \nu(A_j) \right] =\nonumber\\
&&\hspace{3cm} 2 \; D \; \tau \;
\nu(\cM) \; \tensor 1
\ .
\label{sumrule*}
\end{eqnarray}
Equation (\ref{sumrule*}) is fundamental for the following development
because it
constitutes a sum rule relating the diffusion coefficient to the first
coefficients
$\vec{T}(A_j)$ of the wavenumber expansion of the hydrodynamic measures
which is
linear in the wave number $\vk$.  The measures $\vec{T}(A_j)$ have been
interpreted elsewhere as the stationary non-equilibrium measures associated
with a
gradient of concentration of tracer particles across the system.  In the case of
the multi-baker maps, $\vec{T}(A_j)$ is given by the difference of the
Takagi function
at both
ends of the one-dimensional sets $A_j$ \cite{GiDoGa00}.  The sum rule
(\ref{sumrule*}) thus
relates the diffusion coefficient to the generalization of the Takagi
function for
the present system.


\section{Entropy production}

\label{EntrProd}


\subsection{Definitions}


In this section we are going to calculate of the rate of irreversible
entropy production over a time $\tau$, assuming that $t \gg\tau$.

For our calculation of the rate of entropy production in a unit cell of the
periodic lattice, we use a partition of the total
phase space into the small disjoint sets $A_j$ defined above Eq. (\ref{22}).
We suppose that the partition is
invariant under the spatial translations ${\cal T}^{\vl}$.  The phase space
cell located at the lattice vector $\vl$ is decomposed by
this partition as:
\begin{equation}
\cM_{\vl} =\cup_{A_{j}\subset \cM_{\vl}} A_{j}\; .
\label{11}
\end{equation}
This partition can be seen as a translationally invariant grid extending
over the
whole phase space.

We begin by defining the entropy of the lattice cell $\cM_{\vl}$ at time $t$
as the coarse-grained entropy of this cell with respect to the partition
(\ref{11}):
\begin{eqnarray}
S_t(\cM_{\vl}|\{A_j\}) &\equiv& -\sum_{A_j\subset
\cM_{\vl}}\mt(A_j)\; \ln
\frac{\mt(A_j)}{\nu(A_j)} \nonumber\\
&&+ S_{\rm eq}(\cM_{\vl}|\{A_j\}) \ ,
\label{34}
\end{eqnarray}
where we have set Boltzmann's constant equal to unity, $k_{\rm
  B}=1$. The first term on the right hand side of Eq. (\ref{34})
  is the non-equilibrium relative entropy with respect to the
  equilibrium entropy for this partition. The equilibrium entropy is given by
\begin{equation}
S_{\rm eq}(\cM_{\vl}|\{A_j\}) =  -\sum_{A_j\subset
\cM_{\vl}}\nu(A_j)\; \ln{\frac{\nu(A_j)}{c}},
\label{34a}
\end{equation}
where $c$ is a constant which fixes the absolute value of the
equilibrium entropy.
The time variation of the entropy over a time interval $\tau$ is
of course only due to the change in the relative entropy, and is defined
as the
difference
\begin{widetext}
\begin{eqnarray}
\Delta^{\tau}S(\cM_{\vl}) &\equiv& S_t(\cM_{\vl}|\{A_j\}) -
S_{t-\tau}(\cM_{\vl}|\{A_j\}) \ ,\nonumber\\
&=& S_t(\cM_{\vl}|\{A_j\}) -
S_t(\Phi^{\tau}\cM_{\vl}|\{\Phi^{\tau}A_j\})\ .
\end{eqnarray}
On the other hand, the
{\it entropy flow} is defined as the difference between the
entropy which enters the cell $\cM_{\vl}$ and the entropy which exits that
cell:
\begin{eqnarray}
\Delta_{\rm e}^{\tau}S(\cM_{\vl}) &\equiv&
S_{t-\tau}(\Phi^{-\tau}\cM_{\vl}|\{A_j\}) -
S_{t-\tau}(\cM_{\vl}|\{A_j\}) \ ,\nonumber\\
&=& S_t(\cM_{\vl}|\{\Phi^{\tau}A_j\}) -
S_t(\Phi^{\tau}\cM_{\vl}|\{\Phi^{\tau}A_j\})\ .
\end{eqnarray}
\end{widetext}
Accordingly, the
{\it entropy production} over a time $\tau$, assuming that $t \gg\tau$,
is defined as
\begin{eqnarray}
\ds(\cM_{\vl}) &\equiv& \Delta^{\tau}S(\cM_{\vl}) -\Delta_{\rm
e}^{\tau}S(\cM_{\vl})\ , \nonumber\\
&=&  S_t(\cM_{\vl}|\{A_j\}) -
S_t(\cM_{\vl}|\{\Phi^{\tau}A_j\})\ .
\label{33}
\end{eqnarray}


\subsection{Calculation of the entropy production}


Equation (\ref{33}) gives the expression of the entropy production as the
difference between
the entropy with respect to the original partition into sets $A_j$ and the
entropy
with respect to a partition into sets which are the images $\Phi^{\tau}A_j$
of the
sets $A_j$ after time $\tau$.  We notice that  each set
$\Phi^{\tau}A_j$ belongs to a single unit cell $\cM_{\vl}$ by a previous
assumption.  Moreover, since the partition is invariant under translation from
cell to cell, the partition $\{\Phi^{\tau}A_j\}$ is identical to the partition
$\{\phi^{\tau}A_j\}$ obtained by using the flow on the torus.

Written out in full, this entropy production is
\begin{eqnarray}
\ds(\cM_{\vl}) & = &  -\sum_{A_j\subset
\cM_{\vl}}\mt(A_j)\; \ln\frac{\mt(A_j)}{\nu(A_j)}
\nonumber \\ && +
\sum_{\phi^{\tau}
A_j\subset\cM_{\vl}}\mu_t(\phi^{\tau}A_j)\;
\ln\frac{\mu_t(\phi^{\tau}A_j)}{\nu(A_j)}
\label{35}
\end{eqnarray}
where we have used $\nu(\phi^{\tau}A_j)=\nu(A_j)$.
Next we expand in powers of the deviations of the measures from their
equilibrium values and find
\begin{eqnarray}
\ds(\cM_{\vl}) &=&  {1\over
 2}\sum_{\phi^{\tau}
A_j\subset\cM_{\vl}}\frac{\left[\delta\mt(\phi^{\tau}A_j)\right]^{2}}
{\nu(A_j)} \nonumber\\
&&-
{1\over 2}\sum_{A_j\subset\cM_{\vl}}
\frac{\left[\delta\mt(A_j)\right]^{2}}{\nu(A_j)}
+  O\left(\delta\mu_t^{3}\right) \ .
\label{36}
\end{eqnarray}

We now use the explicit forms for the measures $\delta\mt(A)$  given
by the second term on the right hand side of Eq. (\ref{18}). After
some algebra and the use of the conservation of measures, as well as
the summation formulas (\ref{sumT=0}) and (\ref{sumg=0}), we find that the
right
hand side of Eq. (\ref{36}) becomes
\begin{widetext}
\begin{eqnarray}
\ds(\cM_{\vl}) &=& {1\over
2}\; {\frac{1}{|{\cal{B}}|^{2}}}\int_{\cal{B}}\; d\vk_1\;
F_{\vk_1}\int_{\cal{B}}\;d\vk_2 \; F_{\vk_2}
e^{i\vl\cdot(\vk_1
+\vk_2)}\; C(\vk_1,t)\; C(\vk_2,t)\nonumber \\
&&\times  e^{(s_{k_1}+s_{k_2})t}
\sum_{j} \frac{1}{\nu(A_j)}\vk_1\vk_2:\left[\vec{T}(A_j)\vec{T}(A_j)
-\vec{T}(\phi^{\tau}A_j)\vec{T}(\phi^{\tau}A_j)\right]\ .
\label{37}
\end{eqnarray}
Here the summation is over the sets $A_j$ that form a partition of the
unit cell $\cM_{\vl}$. Now we use the identity (\ref{29}) and the sum
rule (\ref{sumrule*}) to obtain our central result
\begin{eqnarray}
\ds(\cM_{\vl}) & = &
-D\; \tau\; {\frac{\nu(\cM)}{|{\cal{B}}|^{2}}}\int_{\cal{B}}\;d\vk_1
\; F_{\vk_1}\int_{\cal{B}}\;d\vk_2 \; F_{\vk_2}\;
\vk_1\cdot\vk_2
e^{i\vl\cdot(\vk_1 +\vk_2)}\; C(\vk_1,t)\; C(\vk_2,t)\;
e^{(s_{k_1}+s_{k_2})t}
\nonumber \\
& \simeq & D\; \tau\; \frac{1}{n_{\rm eq}} \; \left[
\frac{\partial
n(\vl,t)}{\partial\vl}\right]^2\; ,
\label{38}
\end{eqnarray}
\end{widetext}
since the tracer density is expressed according Eq. (\ref{8/7}) and
$n_{\rm eq}=\nu(\cM)$.

Here we have used the isotropy of the motion, in order
to eliminate correlations between the displacements in orthogonal directions.
This results in the factor $\vk_1\cdot\vk_2$
appearing in the integrand in Eq. (\ref{38}). We have also implied a
scaling limit in order to write the last line of this equation.



\section{Conclusions}


\label{Concl}

In this paper, the irreversible entropy production has been derived from
statistical mechanics for a process of diffusion in periodic, deterministic
dynamical
systems.  The derivation starts from Gibbs' coarse-grained entropy,
assumes that the system is spatially periodic, and that the dynamics
satisfies Liouville's
theorem, is chaotic and mixing. We chose a coarse graining
partition of phase space which has the property that any two
trajectories starting in the same set of the partition will remain
close together over some specified time interval.
The central quantities appearing in our
derivation are the hydrodynamic measures. They define, at the microscopic
level, the
hydrodynamic modes of diffusion which are exponentially damped at a rate
given by
the van Hove dispersion relation for diffusion. These
hydrodynamic measures describe the
approach
to the thermodynamic equilibrium under the diffusion process.  In deterministic
systems, the hydrodynamic measures turn out to be singular. Indeed, the
quantity
$\vec{T}(A)$ is a measure describing a non-equilibrium stationary state
corresponding to a gradient of concentration across the system and, in the
multi-baker, the cumulative function of this measure is known to be the
continuous
but non-differentiable Takagi function \cite{TaGa95}.  On the other hand,
it has
been shown elsewhere that the hydrodynamic measures are singular in periodic
Lorentz gases \cite{GaClGiDo01}. It should be emphasized here that the
hyperbolicity of the system is used to argue that the time dependent
hydrodynamic measures, defined by Eq. (\ref{17}) approach
their asymptotic forms, with variations on arbitrarily fines scales in
phase space, on time scales determined by the positive
Lyapunov exponents, which are very short compared to the time scales
necessary for the relaxation of the density distribution to
equilibrium. Of course, not all systems with macroscopic diffusion and a positive
rate of entropy production are chaotic, {\it e.g} wind-tree models with
non-overlapping trees\cite{HaugeCohen}. In such systems the mechanism
responsible for positive entropy production is expected to develop on
algebraic rather than exponential time scales, as is the case for
chaotic systems.

Moreover, if the
hydrodynamic measures were regular, expression (\ref{38}) for the entropy
production would vanish, as shown in the case of the multi-baker map
\cite{Ga97,GiDoGa00}.  Accordingly, the singular character of the
hydrodynamic measures 
plays an essential role in the
positiveness of the entropy production expected from irreversible
thermodynamics.  A further point we want to emphasize is that the present
derivation leads naturally to a positive entropy production in agreement
with the
second law of thermodynamics. Finally we point out that our derivation
of the expression given by irreversible thermodynamics for the rate of
entropy production applies not only to Lorentz gases, but also to
tracer diffusion taking place in a spatially periodic $N$ particle
system, where the mechanism for tracer diffusion is provided by the
particles interacting with each other. Without this interaction, the
motion of the tracer particle would be ballistic, and the mean square
displacement would grow quadratically with time.

The next steps to be taken in the development of this approach to the
theory of
entropy production in the relaxation of fluid systems to
thermal equilibrium is to generalize the method given here for tracer
diffusion in a periodic $N$ particle system to viscous and heat flows.
We shall report on the derivation of entropy production for these other
transport
processes in future publications. Finally one would like to remove the
restriction to periodic systems, and to consider the entropy
production for a general, isolated $N$ particle system relaxing to
thermal equilibrium. This remains open for future work.

It is necessary mention that our approach to entropy production
in
fluids has been criticized by Rondoni and Cohen in a series of
papers\cite{RonCo}. This is not the place to provide a detailed
response to their criticisms, which we will do in separate
publications. However, it is appropriate here to mention two issues
that are of some importance for our response to their comments. (1)
Rondoni and Cohen find it troublesome that the time needed for our
method to be applicable depends on the nature of the partition chosen
for the calculation of the relative entropy and its change with
time. While this observation is indeed correct, the consequences are
not problematic. The essential point to note is that for a non-zero rate
of irreversible
entropy production to result from our method, we require a partition
which is coarser than the scale
of variation of the non-equilibrium distribution function. As we have
tried to make clear, the non-equilibrium distribution function develops
a fractal structure on a time scale set by the magnitude of the largest
positive Lyapunov exponent. For the systems described here, this is a
time scale which is very short compared to the time scale over which
the system relaxes to equilibrium. In fact, this time scale is on the
order of the mean free time between collisions, a scale sufficiently
short that more traditional methods for computing the rate of entropy
production, based upon the Boltzmann equation, for example, do not
apply, either. (2) Rondoni and Cohen have also objected to our use of
multi-baker maps and Lorentz gases as examples of systems to which
non-equilibrium thermodynamics might be applied. While one might
object to their view of the utility of simple model systems for the
development of physical intuition, our
calculation of the rate of entropy production for tracer diffusion in
an $N$ particle system, shows that the method can be applied to much
more general systems than multi-baker maps and Lorentz gases. Our
future work will be devoted to seeing exactly how general this method
might be.

{\bf Acknowledgments:}
The authors would like to thank the Max Planck Institute for 
Physics of Complex Systems, Dresden, for its hospitality while part of
this work was carried out. JRD thanks the National Science
Foundation for support under grant PHY 98-20824;  PG thanks the
National Fund for Scientific Research (FNRS Belgium) for financial support;
TG thanks the Israeli Council for Higher Education and the Feinberg
Postdoctoral Fellowships
Program at the Weizmann Institute of Science for financial support.

\end{document}